\documentclass{kluwer}
\begin{document}
\begin{article}
\begin{opening}
\title{Finding Faint Intermediate-mass Black Holes in the Radio Band}            
\author{T. J. \surname{Maccarone}\email{tjm@science.uva.nl}}
\institute{University of Amsterdam}
\author{R. P. \surname{Fender}} 
\institute{University of Southampton}                               
\author{A. K. \surname{Tzioumis}}
\institute{Australian National Telescope Facility}

%\date: rather not
\runningtitle{Faint IMBHs}
\runningauthor{Maccarone, Fender \& Tzioumis}

\begin{ao}
tjm@science.uva.nl
\end{ao} 

\begin{abstract} 
We discuss the prospects for detecting faint intermediate-mass black
holes, such as those predicted to exist in the cores of globular
clusters and dwarf spheroidal galaxies.  We briefly summarize the
difficulties of stellar dynamical searches, then show that recently
discovered relations between black hole mass, X-ray luminosity and
radio luminosity imply that in most cases, these black holes should be
more easily detected in the radio than in the X-rays.  Finally, we
show upper limits from some radio observations of globular clusters,
and discuss the possibility that the radio source in the core of the
Ursa Minor dwarf spheroidal galaxy might be a $\sim 10,000-100,000
M_\odot$ black hole.
\end{abstract}

%\keywords{}
\end{opening}
\section{Introduction}

Black holes are generally found to exist in two classes - the stellar
mass (i.e. about 10 $M_\odot$) black holes which are usually found as
a result of their being in X-ray binaries, and the galactic mass
(i.e. $10^6-10^9 M_\odot$) black holes found in the cores of massive
galaxies.  Some recent evidence has begun to develop for black holes
with masses just below $10^6 M_\odot$ (e.g. Fillipenko \& Ho 2003;
Greene \& Ho 2004), and some of the ultraluminous X-ray sources show
evidence for spectral (e.g. Jon Miller's contribution in this volume)
and variability (Strohmayer \& Mushotzky 2003) characteristics that
would indicate black holes of 100-1000 $M_\odot$.

Whether globular clusters and dwarf spheroidal galaxies have black
holes in their centers is an especially controversial topic.  Some
authors have argued that globular clusters should build up black holes
of roughly 0.1\% of their total mass through mergers of stellar mass
black holes (Miller \& Hamilton 2002); this fraction of the total mass
is roughly the same as the fraction of galactic bulges' masses locked
up in their central black holes (Magorrian et al. 1998), and there
have been some claims of observational evidence for these
intermediate-mass black holes (see e.g. Gebhardt, Rich \& Ho 2002;
Gerssen et al. 2002).  On the other hand, it has been argued that the
central increases in mass-to-light ratio in globular clusters can be
equally well explained by concentrations of white dwarfs
(e.g. Baumgardt et al. 2003; Gerssen et al. 2003), and that black
holes should be dynamically ejected from globular clusters (Portegies
Zwart \& McMillan 2000).  It has been suggested (Drukier \& Bailyn
2003) that searches for individual high velocity stars could be the
most effective way to search for dynamical evidence of black holes in
globular clusters, but that there might not be enough stars to sample
the gravitational potentials of globular clusters well enough to make
use of this method.

In dwarf spheroidal galaxies, the determination of whether there are
intermediate-mass black holes is at least as wide open.  As the escape
velocities of dwarf spheroidal galaxies are typically smaller than
those of globular clusters, it is more likely that the gravitational
radiation recoil effect should work to eject black holes participating
in unequal mass mergers from these systems (e.g. Favata et al. 2004).
On the other hand, it has been suggested on theoretical grounds that
because the dwarf galaxies are likely to be the first galaxies formed,
they should have the densest peaks in their early-universe densities,
and hence might have formed very high mass black holes during their
Population III phases - black holes with 5-40\% of their stellar mass
(Ricotti \& Ostriker 2004).  Observational constraints on the central
mass concentrations of dwarf spheroidal galaxies are generally quite
poor.

In both the dwarf spheroidal galaxy and globular cluster cases, it
seems reasonable to search for new means of finding these black holes.
Almost thirty years ago, it was suggested that the X-ray emission from
globular clusters might be intermediate-mass black holes accreting
from the interstellar medium (Bahcall \& Ostriker 1975).  The
detection of Type I X-ray bursts from the bright globular cluster
sources has since ruled out this possibility, but much deeper
observations from the Chandra Observatory have placed tight
constraints on the X-ray emission that could be coming from such
sources (Grindlay et al. 2001; Ho, Terashima \& Okajima 2003).  In a
few globular clusters (M~15 and 47~Tuc), variations in the pulsar
dispersion measures provide measurements of the density of the
interstellar medium (Freire et al. 2001), and combining these gas
density measurements with the X-ray upper limits in these globular
clusters still fails to prove that black holes of about 1/1000 of the
total mass of the cluster are not present, if one makes reasonable
assumptions about the fraction of the Bondi-Hoyle rate that is
typically accreted in low luminosity systems (see e.g. Bower et
al. 2003; Perna et al. 2003) and on the radiative efficiency of this
accreted material (see e.g. Narayan \& Yi 1994; Fender, Gallo \&
Jonker 2003).  In light of the new fundamental plane relations for
black hole activity (Merloni, Heinz \& Di Matteo 2003; Falcke,
K\"ording \& Markoff 2004), which show that the radio luminosity,
$L_R$ and the X-ray luminosity $L_X$ of a black hole with mass $M_{BH}$
are related such that $L_X \propto L_R^{0.6} M_{BH}^{0.8}$, it has
been shown that the most efficient way to search for evidence of
accretion from low luminosity, high mass black holes such as those
predicted to exist in globular clusters and dwarf spheroidal galaxies
is by searching for radio emission (Maccarone 2004).  For example, a
single 12-hour observation of M~15 with the VLA would place a better
constraint on the existence of a $1000~M_\odot$ black hole than would
the entire mission lifetime of the Chandra Observatory.  The prospects
for improving the sensitivity limits in the radio in the near future
are excellent; the VLA expansion project will improve its sensitivity
by a factor of about 10, making $\mu$Jy level observations possible in
quite short exposure times; the High Sensitivity Array is already
allowing $\mu$Jy level VLBI scale interferometry, and the Square
Kilometer Array project provides some hope that these (and better)
sensitivity levels will be reachable from the Southern Hemisphere
within the next 20 years.

In this contribution, we will outline the basic method by which we
attempt to predict the radio fluxes from intermediate-mass black holes
in globular clusters, and then we will describe some progress that has
been made towards this goal.

\section{Methodology}

In Maccarone (2004), a methodology for going from a globular cluster's
mass and distance to its expected radio flux was laid out.  In this
paper, we will briefly summarize the assumptions of that work, but we
refer the reader to Maccarone (2004) and to the other cited references
for a full justification of each assumption.  Specifically, we assume:

\begin{itemize}
\item A black hole mass of 0.1\% of the globular cluster's stellar
mass (Miller \& Hamilton 2002)
\item A gas density of 0.15 H cm$^{-3}$, approximately the value estimated from pulsar dispersion measures in M~15 and 47~Tuc, and expected from stellar mass loss (Freire et al. 2001).
\item Accretion at 0.1-1\% of the Bondi rate, with the sub-Bondi rate
due to disk winds and/or convection as constrained by observations of
low luminosity AGN in the Galactic Center and in elliptical galaxies
and the lack of observations of isolated neutron stars accreting from
the interstellar medium (e.g. Bower et al. 2003; Perna et al. 2003).
\item A radiative efficiency in the X-rays of:
\begin{equation}
\eta=(0.1)\times\left(1+\frac{A^2}{2 L_{tot}}-A\sqrt{\frac{A^2}{4 L_{tot}^2}+\frac{1}{L_{tot}}}\right),
\end{equation}
with $A$ a constant to be fitted from observations (and being larger
when the jet's kinetic power is a larger fraction of the total
accretion power), and $L_{tot}$ the radiative plus kinetic luminosity
of the system in Eddington units, as used by Fender, Gallo \& Jonker
(2003) to explain the $L_X-L_R$ correlation observed by Gallo, Fender
\& Pooley (2003), with the idea being that enough kinetic power is
pumped into a jet to make the accretion flow radiatively inefficient
(see also, e.g. Malzac, Merloni \& Fabian 2004).  Whether the
radiative inefficiency is due to mass and energy loss into a jet or
also partially due to advection into the black hole (e.g. Ichimaru
1977; Narayan \& Yi 1994) is not clearly established by these
relations, and does not affect the results presented here.  We have
set $A=6\times10^{-3}$ for these calculations, which is based on a
conservative estimate of the jet power.  This relation is used to
convert a calculated accretion rate (from the assumed fraction of the
Bondi-Hoyle rate) into a calculated X-ray luminosity.
\item The fundamental plane relationship among X-ray luminosity, radio
luminosity and black hole mass of Merloni et al. (2003), parameterized
for convenient applications to Galactic globular clusters:
\begin{equation}
F_{5 GHz} = 10 \left(\frac{L_x}{3\times10^{31} {\rm ergs/sec}}\right)^{0.6} 
\left(\frac{M_{BH}}{100 M_\odot}\right)^{0.78} \left(\frac{d}{10 {\rm kpc}}\right)^{-2} {\rm {\mu}Jy}.
\label{merloni}
\end{equation}
This relation is used to convert the calculated X-ray luminosity from
the previous step into a radio flux.  We note that the relation found
by Falcke et al. (2004), which considered only flat spectrum, low
luminosity radio sources like those we expect to see in the centers of
dwarf galaxies or globular clusters, is consistent with this relation
within the uncertainties.  Following these assumptions, several
globular clusters should have central radio sources brighter than a
few $\mu$Jy, but only Omega Cen should have a radio source brighter
than 40 $\mu$Jy, and even Omega Cen should be that bright only if the
accretion rate is closer to 1\% of the Bondi-Hoyle rate than it is to
0.1\% of the Bondi-Hoyle rate.  Nonetheless, the predicted X-ray
fluxes for the globular clusters are generally well below
detectability levels, even with very long observations by the Chandra
Observatory.
\end{itemize}

\section{Applications to Globular Clusters}

We have applied this method to two globular clusters so far, and in
both cases have found no evidence for an accreting central black hole.
Omega Cen was observed by us with the Australian Telescope Compact
Array (ATCA) simultaneously at 4.8 and 8.6 GHz for 12 hours on
8~May~2004, with no detection made at either frequency.  The
non-detection yielded a 3$\sigma$ upper limit on the flux of just
under 100 $\mu$Jy, under the assumption that the radio spectrum of the
source should be flat.  Maccarone (2004) predicted that the flux level
should be at least 150 $\mu$Jy under the most conservative set of
parameter values used in that paper.  This would seem to imply that
there cannot be a black hole with 0.1\% of the cluster's mass in Omega
Cen, but given the scatter in the fundamental plane relation, and the
fact that the gas density in Omega Cen has been assumed to be similar
to those in M~15 and 47~Tuc, rather than measured, this upper limit
should not be interpreted as such strong evidence against an
intermediate-mass black hole.  The upper limit does, however, provide
strong evidence against the combination of 0.1\% of the cluster mass
being in an intermediate black hole, with an accretion rate of
$\sim1$\% or more of the Bondi rate.

We have also considered previous radio observations of M~15 which were
made with the Very Large Array (VLA) for the purposes of finding radio
pulsars (Johnston, Kulkarni \& Goss 1991).  These data were taken at
1.4 GHz, and reached a noise level of 43 $\mu$Jy, with no unidentified
sources found within the core of the globular cluster.  The upper
limits are thus rather similar to those found for Omega Cen, in terms
of flux level.  The constraints on whether there exists a black hole
with 1/1000 of the cluster mass, though, are much weaker, because the
cluster is smaller and further away than Omega Cen.  A useful
constraint can be made on whether there exists a black hole
substantially more massive than this.  In the context of the
assumptions listed above, the $3\sigma$ upper limit for the radio flux
corresponds to the flux level expected from a 700 $M_\odot$ black hole
accreting 0.1\% of its Bondi rate - therefore, the upper limit on the
radio flux measured in this cluster's core can be taken as evidence
against the claimed 2500 $M_\odot$ black hole in M~15 (Gerssen et
al. 2002), although it should be noted that the uncertainty on this
mass measurement was rather large, and the measurement was not
inconsistent with a black hole of 700 $M_\odot$. Proposed High
Sensitivity Array observations could reduce the noise level in the
radio data by a factor of about 10, which could, in turn allow for
either a detection of the black hole or a truly constraining upper
limit on its possible mass.

\section{Applications to Dwarf Spheroidal Galaxies}

We have also searched the NRAO VLA Sky Survey (NVSS) catalog around
the centers of the Milky Way's Northern Hempisphere dwarf spheroidal
galaxies.  This catalog has some sources as faint as 1 mJy, but is
complete only at the level of about 3-4 mJy, and it covers the entire
sky north of a declination of -40 degrees at a frequency of 1.4 GHz
(Condon et al. 1998).  One source was found within the 3$\sigma$ error
circle of the center of a dwarf spheroidal galaxy - a 7.1 mJy source
about 20'' from the reported center of the Ursa Minor dwarf spheroidal
galaxy.  The Ursa Minor galaxy is one of the nearest ($d$=66 kpc),
most diffuse (the 20'' offset is roughly the 1$\sigma$ error in the
centroid position of the galaxy), and most massive ($M=2.3\times10^7
M_\odot$) of the Milky Way's dwarf spheroidal satellites (see Mateo
1998 for a review of the properties of dwarf spheroidal galaxies
including measurements of parameter values).  The density of NVSS
sources on the sky is such that there is about a 5\% chance of finding
a source within the 3$\sigma$ error box of the center of the Ursa
Minor dwarf spheroidal galaxy.  The other dwarf galaxies are further
away and more centrally concentrated, so their centroid positions are
more well established and the chance of a spurious coincidence between
a radio source and their core positions would be quite small.  On the
other hand, because they are further away and less massive, their
expected radio fluxes would be smaller than that of the Ursa Minor's
core, if one assumes there should be a linear correlation between
galaxy mass and black hole mass.  The measurements of gas contents of
dwarf spheroidal galaxies are mostly upper limits (although see
Bouchard, Carignan \& Mashchenko 2003 for one detection), so it is not
as straightforward to convert a radio flux into a black hole mass as
it would be in a globular cluster.  If we assume that the gas density
is 1/30 to 1/100 as high in dwarf spheroidal galaxies as in globular
clusters (because the dwarf spheroidals are more diffuse), then we
find a black hole mass of about 1-2$\times10^5 M_\odot$ would be
required to produce the observed radio flux.  The expected X-ray
luminosity from such an object would be $\sim$ a few $\times10^{34}$
ergs/sec, below the detection limits of past X-ray observations
(e.g. Markert \& Donahue 1985; Zang \& Meurs 2001), but easily
detectable by Chandra or XMM.  Because the error circle of the NVSS
source is about 4'' in radius, it is not possible to identify a unique
optical counterpart and determine whether this radio source is more
likely to be in the Ursa Minor galaxy or a background AGN.  Follow-ups
in radio and X-ray have been proposed, both to obtain a better
positional accuracy for the radio source, and to determine its X-ray
to radio flux ratio.

\section{Prospects for Future Improvements}

One key area for future improvements of this work is to get deeper
radio observations of the globular clusters and dwarf spheroidal
galaxies most likely to show radio sources associated with
intermediate-mass black holes.  This work is already in progress, with
an application in submission for High Sensitivity Array time to
observe M~15.  Unfortunately, most of the best globular cluster
candidates are in the Southern Hemisphere, and with the ATCA data
showing only upper limits for Omega Cen, the prospects of detecting a
black hole in any other globular cluster by using ATCA seem remote.

The other key area that needs more work is in improving our
measurements of the gas densities in these systems, especially in the
dwarf spheroidal galaxies.  The methodology for doing so is not as
clear.  Searches for absorption lines in the spectra of background AGN
seem to be one of the most promising routes (Tinney, Da Costa \&
Zinnecker 1997), and such AGN should be detected as part of any
program searching for X-ray emission from a central black hole as
well.

\acknowledgements It is a pleasure to thank the following for useful
discussions which contributed to this work: Dave Meier, Heino Falcke,
Eva Grebel, Russell Edwards, Andrea Merloni, Cole Miller, Simon
Portegies Zwart, Fred Rasio, Ben Stappers and Kathy Vivas.

\end{article}
\end{document}